\documentstyle[12pt]{article}
\begin{document}
\title{ONE-PARAMETRIC FAMILY \\
OF THE DOUBLE-SCALING LIMITS \\ IN THE HERMITIAN MATRIX MODEL $\Phi^6
:$\\ ONSET OF NONDISSIPATIVE SHOCK WAVES}
\author{ Vadim R. KUDASHEV and Bulat I. SULEIMANOV
\\ Institute of Mathematics, Russian Academy of Sciences,
\\ Chernyshevsky Street 112, 450000 Ufa,
Russian Federation \\ e-mail: kudashev@imat.rb.ru, bis@imat.rb.ru }
\date{\null}
\maketitle
\vskip -1em

 We construct a one-parametric family of the
 double-scaling limits in the hermitian matrix
 model $\Phi^6$ for $2D$ quantum gravity. The
 known limit of Bresin, Marinari and Parisi
 \cite{bre} belongs to this family. The family is
 represented by the Gurevich-Pitaevskii solution of
 the Korteveg-de Vries equation which describes
 the onset of nondissipative shock waves in media
 with small dispersion. Numerical simulation of
 the universal Gurevich-Pitaevskii solution is
 made.

\vspace{1em}
1.In this paper we construct a one-parametric family of the
double-scaling limits in the hermitian matrix model $\Phi^6$
for two-dimensional quantum gravity. This family is
described by a common solution of the Korteweg-de Vries
(KdV) equation
\begin{equation}
v_{t}+vv_{x}+v_{{xxx}}=0
\label{kdv}
\end{equation}
and the following ordinary differential equation (ODE)
\begin{equation}
v_{{xxxx}}+5vv_{{xx}}/3+5(v_{x})^2/6+ 5(x-tv+v^3)/18=0.
\label{one}
\end{equation}
The leading term of the asymtotics of this solution as $|x|$$\to$$\infty$
is
the solution of the cubic equation
\begin{equation}
 x-tf+f^3=0.
\label{cusp}
\end{equation}
In the particular case $t=0$ the solution of (\ref{one}),(\ref{cusp}) is
reduced to the known solution of \cite{bre}. On the other hand, the common
solution of (\ref{kdv}) and (\ref{one}) exactly coincides \cite{bis} with the
well known Gurevich-Pitaevskii (G-P) special solution of the KdV equation
\cite{gur}, \cite{pit} which universally describes the onset of oscillating
dissipationless shock waves in media with small dispersion.

2.An essential development in the theory of two-dimentional
quantum gravity based on the study of $n$$\times$$n$
hermitian matrix models
\begin{equation}
Z_{n}=\int dHexp(-\beta TrU(H)),
\label{her}
\end{equation}
where $H$-$n$$\times$$n$-hermitian matrix,
$U(z)=\sum_{1}^{N} g_{j}z^{j},$
was made in the series of works \cite{kaz}--\cite{ban}. The progress was
achieved, considering the double-scaling limits in models (\ref{her}),
($h$$\to{0}$):
$\beta$=$h^{-4-2/m}B$, $n/\beta$=$(1+\delta x h^{4})A.$
The following remarkable circumstance was then observed \cite{mig},\cite{she}:

It turns out \cite{mig},\cite{she} that the calculation of
the second derivative of the limit of the non-regular part
of $\ln$ $Z_{n}$ ( its regular part is irrelevant) is
reduced to the finding of the limit of solutions $R_{n}$
of nonlinear difference equations
\begin{equation}
n/\beta=Q(R_{n},R_{n+1},R_{n-1},...,R_{n+l},R_{n-l}),
\label{ana}
\end{equation}
which the right hand sides $Q$ uniquely defined by the potentials $U(z)$.
It was established that for any $U(z)$
there exist a
natural $m>1$ and constants $A$, $B$ and $\delta$, such that the asymptotics
of solutions $R_{n}$ of (\ref{ana}) is described by the formula ($\rho$ is a
constant):

$$R_{n}=\rho(1+h^{4/m}v(x)+...)/3,$$

where functions v(x) satisfy of the first Painleve ODE ( for $m=2$ ) and its
higher analogues ( for $m>2$).
Before passing to a detailed discussion of a particular case of the potential
\begin{equation}
U(z)=z^2/2+g_{1}z^4+g_{2}z^6,
\label{fis}
\end{equation}
which we consider in this article, we would like to give
the following useful general statement which should be
taken into account when calculating double-scaling limits:

For any integer  $k$ the functions $v(n+k)$ are expanded into Tailor series
\begin{equation}
 v(n+k)=v(x+hkÓ)=v(x)+\sum_{l=1}^{\infty}(hkÓ)^{l}v^{(l)}(x)/(l!),
\label{tay}
\end{equation}
in which the constant $c$ does not depend on the set of $g_{j}$ of
(\ref{her}),  the natural $m$, and the constants
 $A$,$B$,$\beta$,$\delta$,$\rho$.

Remark. To make the results of the present paper compatible
with \cite{gur} and \cite{pit}, we use notations different
from \cite{bre}. This is the reason why we fix the following
particular value of the constant $c$: $c=\sqrt{6}$.

3.The equation (\ref{ana}) for the potential (\ref{fis}) has
the form \cite{kit}:
\begin{equation}
n/\beta=R_{n}(1+4J_{1}(n)+6J_{2}),
\label{con}
\end{equation}
$$J_{1}(n)=R_{{n+1}}+R_{n}+R_{{n-1}},J_{2}(n)=(J_{1}(n))^2+R_{{n+2}}R_{{n+1}}-
R_{{n+1}}R_{{n-1}}+R_{{n-1}}R_{{n-2}}.$$
In the case of the
general position , one has the case $m=2$. Assuming that the
constants $\delta$, $\rho$,$A$,$g_{1}$ and $g_{2}$ do not
depend of h, the substitution of
\begin{equation}
n/\beta=A(1+\delta h^4x), R_{n}=\rho(1+h^2v(x)+...)/3
\label{two}
\end{equation}
into (\ref{ana}) and equating the coefficients at different powers of $h$ to
the zero gives the following sequence of relations  \cite{kit}:
\begin{equation}
h^{0}:\rho(60{\rho}^{2} g_{2}+12\rho g_{1}+1)=A,
\label{nil}
\end{equation}
\begin{equation}
h^{2}: {\rho}^{2}(240\rho g_{2}+24g_{1})=-2A,
\label{nex}
\end{equation}
\begin{equation}
h^{4}:-(g_{1}+10\rho g_{2}) \delta x=(15\rho g_{2}+g_{1})(2v_{xx}+v^{2}).
\label{per}
\end{equation}
It follows from the first two relations that $\rho$ is a solution of the
quadratic equation
\begin{equation}
180g_{2}{\rho}^2+24g_{1}\rho+1=0.
\label{squ}
\end{equation}
Except for the degenerate case
\begin{equation}
g_{2}=-g_{1}/(15\rho)=4g_{1}^{2}/5,
\label{sin}
\end{equation}
(i.e. the case of the multiplicity of roots in $(\ref{squ})$),  the function
$v(x)$ turns out to be a solution of the first Painleve equation.

In the degenerated case, one needs another limiting transition. Assuming that
(\ref{sin}) is true, Bresin, Marinari and Parisi considered in \cite{bre} the
following limiting transition:
\begin{equation}
R_{n}=-(1+h^{2}v(x))/12(g_{1}),n/\beta=-(1-h^{6}x)/36(g_{1}),
\label{vyr}
\end{equation}
where $v$ is solution of ODE
\begin{equation}
v_{xxxx}+5vv_{xx}/3+5(v_{x})^2/6+5(x+v^3)/18=0,
\label{mar}
\end{equation}
\begin{equation}
v(x)\sim -x^{1/3}, |x|\to\infty.
\label{bou}
\end{equation}
The numerical solution of the boundary value problem (\ref{mar}),(\ref{bou})
was obtained in \cite{bre} ( it was based on a difference scheme in which the
ODE (\ref{mar}) was replaced by the starting discrete equation (\ref{con}).)
That numerical calculation has shown the uniqueness of the solution of the
boundary value problem (\ref{mar}), (\ref{bou}).

The work \cite{bre} had a wide response. The limiting transition considered in
that paper was studied later from different points of view in a series of
articles (see e.g. example \cite{kit},\cite{moo},
). An unexpected
connection of that transition with the an old problem of the onset of
dissipationless shock waves was discovered in 1994 in \cite{bis}. It was
observed that the solution of the boundary value problem (\ref{mar}) coincided
with the G-P special solution of the equation (\ref{kdv}) at $t=0$.

However, in spite of the significance of \cite{bre}, the
given analysis of the limiting transition in the degenerated
case was not satisfactory from the view point of a general
requirement for investigations of degenerate cases of that
kind. According to the same requerment going back to
H.Puancare, "... the investigation of degenerate systems
should not be restricted by the study  of the picture in the
point of degeneration, but should include the description of
the reorganizations which take place when the parameter
passes through the degenerated value" \cite{arn}.

It will be shown in the next section if we try to satisfy
that requirement and take into account the additional
statement (\ref{tay}),  we are led to the replacement of
the boundary value problem (\ref{mar}),(\ref{bou}) by the
one-parametric family of boundary value problems
(\ref{one}),(\ref{cusp}). The particular case of the
double-scaling limit in the degenerate case (\ref{sin})
exactly corresponds to $t=0$.

4. Assume  that the critical difference $g_{1}-4g_{2}^2/5$ has the order
 $O(h^{p})$:
\begin{equation}
g_{1}=a+bh^{p}+...,   g_{2}=4a^2/5+rbh^{p}+... .
\label{poi}
\end{equation}
(Here $p$,$a$,$b$ and $r$ are constants independent on $h$.) It follows from
the square equation (\ref{squ}) follows that in this case
\begin{equation}
\rho=-1/(12a)+O(b^{1/2}h^{p/2}),
\label{pop}
\end{equation}
and the relation (\ref{per}) implies that the
parameter $\delta$ should also be small:
\begin{equation}
\label{smo}
\delta=\gamma h^{q}+...
\end{equation}
($q$,$\gamma$ are constants independent on $h$). Assuming
as before that (\ref{nil}), (\ref{nex}) are true,
substituting  the relations (\ref{tay}),
(\ref{two}), (\ref{poi})-(\ref{smo}) into (\ref{con}), and then equating to
zero the coefficiets at $h^{i}$, we autoumatically (upon excluding of
irrelevant trivial cases) come to the conclusion that
\begin{equation}
p=4,q=2,
g_{1}=a(1-th^{4}/3+...), g_{2}=4a^{2}(1-th^4+...)/5.
\label{uto}
\end{equation}
The parameter $t$
is calibrated in accordance with
notations of \cite{pit}.

Taking into account the relations
(\ref{two}),
(\ref{pop}) and (\ref{smo}), we obtain after some changes  of notation  a
one-parametric family of double-scaling limits
\begin{equation}
n/\beta=-(1-h^{6}x+...)/(36a), R_{n}=-(1+h^{2}v(t,x)+...)/(12a),
\label{fre}
\end{equation}
where $v$ is a solution of the ODE (\ref{one}).(The last
statement is obtained if one substitutes
(\ref{tay}),(\ref{uto}),(\ref{fre}) into the discrete string
equation (\ref{con}) and equates coefficient at $h^{6}$ to
zero.)

5. The solution $v(x)$ of the ODE (\ref{mar}) of \cite{bre} satisfies the
boundary conditions (\ref{bou}). On the other hand, the ODE (\ref{one}) is
isomonodromic \cite{bis}, and the complete set of its monodromic data is
completely defined by the boundary condition (\ref{bou}).( It has been
calculated at $t=0$ by Moore in \cite{moo}. ) This implies (see \cite{bis})
that the solution $v(t,x)$ of the ODE (\ref{one}) is at the same time the
solution of the KdV equation (\ref{kdv}), and that the leading term of its
asymptotics as $|x|$ $\to$ $\infty$ is the solution of the cubic equation
(\ref{cusp}). According to \cite{gur}, exactly this feature defines a special
solution of the KdV equation which describes in a general way the onset of
dissipationless shock waves in problems with small dispersion.

Actually, the corresponding solution (\ref{cusp}) is the
leading term of the asymptotics of $v(t,x)$ as $t$ $\to$
$-\infty$  for any $x$ as well. If $t$ $\to$ $\infty$,
this asymptotics is true for any $x$ except for a limited
(but expanding with the growth of $t$) area filled with
high frequency oscillations which correspond to the
process of dissipationless shock wave generation.

For the description of the behavior of $v(t,x)$ in this area it was suggested
in \cite {pit} to use the self-similar solutions of the Whithem equations
\cite{roy}
which arise after averaging with respect to the period of a "knoidal" wave
(one-phase periodic solution of (\ref{kdv}))
The self-similar substitution suggested by Gurevich and
Pitaevskii
(correlating to the self-similarity of the solution of
(\ref{cusp})) reduces the Whithem equations to the ODE.
which was numerically investigated in \cite{pit}. Later a
corresponding solution of the ODE
was found in
an explicit form by Potemin \cite{pon}.
( This solution of Potemin, as has been pointed in \cite{bis}, may be easily
found, using the results of \cite{sha} and the fact that the ODE (\ref{one})
is valid for $v(t,x)$. ) Potemin also has found the exact values of the
so-called trailing $s_{-}$ and leading $s_{+}$ fronts, which define the
boundary of the oscillation region ($s=x/t^{3/2}$): h
$$s_{-}=-\sqrt{2}, s_{+}=\sqrt{10}/27$$
numerically established in \cite{pit} before.

The next step to clarify the behaviour of $v(t,x)$ was done in the recent
paper \cite{kud} in which the uniform behavior of the GP solution in the
neighborhood of the trailing edge was studied (the answer is obtained in the
terms of the separatrix solution of the second Painleve solution). Besides,
the speculations of \cite{kud} show that the "average" description of
\cite{pit} and the real leading term of G-P solution asymptotics are probably
the same within the oscillation region up to the shift of phase of $\pi/2$.
6.However, those results concern only to the behavior of $v(t,x)$ with great
$t$. Taking into account the universal nature of the G-P special solution, the
problem of numerical simulation of the G-P solution's behavior with a finite
$|t|$ (solved in the last section of this article) is of interest as well. To
solve this problem, we use an elementary iterative scheme similar to one of
\cite{bre} in which ODE (\ref{one}) is replaced by a discrete equation
equivalent to the discrete string equation (\ref{con}). Namely, the following
procedure is used ($[- L,L]$ is a numerically investigated interval,
$\epsilon= L/N$ is the step of the uniform net):
$$v_{n+1}(k)=v_{n}(k)+P[v_{n}(k),k],$$
$$P[v(k),k]=R_{k}[1+4g_{1}J_{1}(k)+6g_{2}J_{2}(k)]-(1-{\epsilon}^7k\sqrt{6})/3,$$
$$J_{1}(k)=R_{k+1}+R_{k}+R_{k-1},J_{2}(k)=(J_{1}(k))^2+
R_{k+2}R_{k+1}-R_{k+1}R_{k-1}+R_{k-1}R_{k-2},$$
$$g_{1}=-(1-t{\epsilon}^4/3)/12,g_{2}=(1-t{\epsilon}^4)/180,
R_{k}=(1+{\epsilon}^2v_{k}),$$
$$v_{0}(0)=0, v_{0}(k)=-{({\epsilon}k\sqrt{6})}^{1/3}-
t{({\epsilon}k\sqrt{6})}^{-1/3}/3 (k\not=0)$$ and in all
approximations for $|k|>N-1$:
$v_{n}(k)$=$v_{0}(k).$
Results of the numerical analysis
$1,2$
correspond to the
results of \cite{pit} on the qualitative level. In particular, a good
correspondence has been observed between theoretical results and the positions
of the trailing $s_{-}$ and the leading $s_{+}$ fronts of the oscillation
region.

Authors are grateful to V.E. Adler for the help in the
numerical simulations. This work has been supported by the
Russian Foundation of Fundamental Researches
$(96-01-00382)$. The second author was maintained by
Foundation for Leading Scientific Schools of Russia (
96-15-96241).

\end{document}